\documentstyle{article}
\newtheorem{rem}{Remark}
\newcommand{\br}{\begin{rem}}
\newcommand{\er}{\end{rem}}
\newtheorem{ex}{Example}
\newcommand{\bex}{\begin{ex}}
\newcommand{\eex}{\end{ex}}
\newtheorem{Def}{Definition}
\newcommand{\bd}{\begin{Def}}
\newcommand{\ed}{\end{Def}}
\newtheorem{theorem}{Theorem}
\newcommand{\bt}{\begin{theorem}}
\newcommand{\et}{\end{theorem}}
\newtheorem{lemma}{Lemma}
\newcommand{\bl}{\begin{lemma}}
\newcommand{\el}{\end{lemma}}
\newtheorem{cor}{Corollary}
\newcommand{\bc}{\begin{cor}}
\newcommand{\ec}{\end{cor}}
\newcommand{\be}{\begin{equation}}
\newcommand{\ee}{\end{equation}}
\newcommand{\bea}{\begin{eqnarray}}
\newcommand{\eea}{\end{eqnarray}}
\newcommand{\pa}{\partial}
\newcommand{\nn}{\nonumber}

\newcommand{\adots}{\mathinner{\mkern2mu\raise1pt\hbox{.}\mkern2mu
\raise4pt\hbox{.}\mkern2mu\raise7pt\hbox{.}\mkern1mu}}

\title{Integrable Quartic Potentials and Coupled KdV Equations}
\author{S. Baker${}\sp a$, V.Z. Enolskii${}\sp b$ and A.P. Fordy${}\sp a$}

\begin{document}

\maketitle
{\it
${}\sp a$
Department of Applied Mathematical Studies and
Centre for Nonlinear Studies, University of Leeds, Leeds LS2 9JT, UK \\
e-mail  amt6apf@leeds.ac.uk

${}\sp b$
Department of Theoretical Physics, Institute of Metal Physics,
ul. Vernadsky 36, Kiev-680, 252142, Ukraine.
}

\begin{abstract}
We show a surprising connection between known integrable Hamiltonian systems
with quartic potential and the stationary flows of some coupled KdV systems
related to fourth order Lax operators.  In particular, we present a
connection between the Hirota-Satsuma coupled KdV system and (a
generalisation of) the $1:6:1$ integrable case quartic potential.  A
generalisation of the $1:6:8$ case is similarly related to a different (but
gauge related) fourth order Lax operator.

We exploit this connection to derive a Lax representation for each of these
integrable systems.  In this context a canonical transformation is
derived through a gauge transformation.
\end{abstract}

\section{Introduction}

In \cite{f91-1} a surprising connection was shown between integrable cases of
the H\'enon-Heiles system and the stationary flows of some known integrable
PDEs : the $5th$ order KdV, Sawada-Kotera and Kaup-Kupershmidt equations,
which are integrable through second and third order differential Lax
operators.  This gave rise to a matrix spectral problem for each of the
integrable H\'enon-Heiles systems.  Thus, integrable cubic potentials are
associated with second and third order matrix Lax operators.

In \cite{f91-1} the starting point was the H\'enon-Heiles system, using some
elementary, ad hoc calculations.  An alternative approach is to start with
the stationary flow of a known integrable PDE and to find the appropriate
H\'enon-Heiles coordinates.  The most natural way of doing this is through
the Hamiltonian structure of the PDE.  As motivation, this is presented for
the H\'enon-Heiles system in section 2.

In this note we exhibit a similar connection between some stationary flows
associated with fourth order Lax operators and generalisations of
some integrable Hamiltonian systems with quartic potentials:
\be
H = \frac{1}{2} \left( p_1^2 + p_2^2 \right) +
                 a q_1^4 + b q_1^2 q_2^2 + c q_2^4 .  \label{quartic}
\ee
There are 4 nontrivial cases which are integrable:
\begin{enumerate}
\item $a : b : c = 1 : 2 : 1$ ,   \label{121}
\item $a : b : c = 1 : 12 : 16$ , \label{11216}
\item $a : b : c = 1 : 6 : 1$ ,   \label{161}
\item $a : b : c = 1 : 6 : 8$ .  \label{168}
\end{enumerate}
Various inverse square terms can be added without destroying complete
integrability.  Cases (\ref{121}), (\ref{11216}) and (\ref{161}) are
separable in respectively polar,  parabolic and Cartesian coordinates.
Case (\ref{11216}) is one of the higher degree polynomial examples given in
\cite{93-5}.   Case (\ref{168}) is much more complicated and has only
recently been shown to be separable (in a generalised sense) \cite{94-1},
using a Painlev\'e expansion.

Our main result is to relate cases (\ref{161}) and (\ref{168}) to some
coupled KdV equations associated with fourth order Lax operators, which are
themselves related through a Miura map.  The connection with integrable PDEs
gives a Lax representation for these finite dimensional systems, together
with similarity and canonical transformation, thus giving an explanation of
the separation coordinates found in \cite{94-1}.

\section{Stationary Flows and Integrable H\'enon-Heiles Systems}

It was shown in \cite{f91-1} that the general (not necessarily integrable)
H\'enon-Heiles system can be related to the stationary flow of:
\be
u_t = \left( \pa^3 + 8 a u \pa + 4 a u_x \right) \delta_u
      \left( - \frac{1}{3} b u^3 - \frac{1}{2} u_x^2 \right) .
                                        \label{kdv}
\ee
For this stationary flow, the gradient of the above Hamiltonian is in the
kernel of the third order Hamiltonian structure.  Thus we may write:
$$
\delta_u \left( -\frac{1}{3} b u^3 - \frac{1}{2} u_x^2 \right) = A,
                  \quad \hbox{where} \quad
    \left( \pa^3 + 8 a u \pa + 4 a u_x \right) A = 0 .
$$
We now set $A = \alpha y^2$ to find:
$$
y ( y_{xx} + 2 a u y )_x + 3 y_x ( y_{xx} + 2 a u y ) = 0.
$$
Setting $y_{xx} + 2 a u y = \gamma$, we solve the first order equation for
$\gamma$ to get $\gamma = 2 k y^{-3}$, where $k$ is a constant.  We now
have:
\bea
&& \delta_u \left( -\frac{1}{3} b u^3 - \frac{1}{2} u_x^2 \right) =
                                          \alpha y^2 , \nn  \\
&&  y_{xx} + 2 a u y = 2 k y^{-3} ,               \nn
\eea
which, for $\alpha = - a$, are Lagrangian, with:
$$
{\cal L} = \frac{1}{2} \left( u_x^2 + y_x^2 \right) +\frac{1}{3} b u^3
             - a u y^2 - k y^{-2}.
$$
The standard Legendre transformation now renders a natural Hamiltonian
system, which is just the usual generalisation of the
H\'enon-Heiles system:
\bea
&&  q_1 = u, \quad p_1 = u_x, \quad q_2 = y, \quad p_2 = y_x ,  \nn  \\
&&  h = \frac{1}{2} \left( p_1^2 + p_2^2 \right)
         - \frac{1}{3} b q_1^3 + a q_1 q_2^2 + k q_2^{-2}.   \nn
\eea
Thus the Hamiltonian structure of (\ref{kdv}) gave us a natural way of
defining some interesting coordinates, giving us the H\'enon-Heiles
representation of the stationary flow.  For the integrable cases it is
possible to use the Lax representation of (\ref{kdv}) to derive a matrix Lax
representation for the corresponding H\'enon-Heiles system \cite{f91-1}, thus
proving the complete integrability of the latter.  The gauge equivalence of
the Sawada-Kotera and Kaup-Kupershmidt equations leads to a canonical
transformation between the corresponding H\'enon-Heiles systems ($a/b = - 1/6
\; \; \hbox{and} \;\; a/b = - 1/16$).

In this paper, we obtain similar results for some quartic potentials.

\section{Fourth Order Operators}

We start with the self adjoint fourth order operator, which we write in
factorised form:
\be                         \label{sym}
L_1 =  ( \pa + v_1 ) (\pa + v_2 ) ( \pa - v_2 ) (\pa - v_1 )  .
\ee
This can be written as the product of two second order operators:
$$
L_1  = (\pa^{2} + f \pa + f_{x} + g ) (\pa^{2} - f \pa + g ) ,
$$
to give the Miura map:
\be
f = v_{1} + v_{2}, \quad   g = v_{1} v_{2} - v_{1x}.  \label{miura1}
\ee
The Lax equation:
$$
L_{1t} = [ M_1 , L_1 ]
$$
where:
$$
M_{1} =  2 \pa^{3} + \frac{3}{2} ( 2 g - f_{x} - f^{2} ) \pa
         + \frac{3}{4} ( 2 g_{x} - f_{xx} - 2 f f_{x} ) ,
$$
gives the following coupled KdV system:
\be                    \label{ckdv}
        \begin{array}{lll}
f_{t} &= & -\frac{1}{2} ( 2 f_{xxx} +  3 f f_{xx} + 3 f_{x}^{2}
        - 3 f^{2} f_{x} + 6 f g_{x} + 6 g f_{x}), \\
        && \\
g_{t} &= & \frac{1}{4} ( 2 g_{xxx} + 12 g g_{x}+ 6 f g_{xx}+ 12 g f_{xx}
+ 18f_{x}g_{x}   -6f^{2}g_{x}  \\
& & + 3 f_{xxxxx}  + 3 f f_{xxx} + 18 f_{x} f_{xx} - 6 f^{2} f_{xx}
                -6 f f_{x}^{2} ).
                \end{array}
\ee
\br
The resulting coupled KdV system is simpler in co-ordinates $r$ and $s$,
corresponding to the operator:
$$
L_{sym} =  \pa^4 + \pa r \pa + s ,
$$
but the Miura map and Hamiltonian operator are more complicated, so not
well suited to our purposes.
\er

A rotation of factors in (\ref{sym}) leads to another operator:
$$
L_{2} = ( \pa - v_{1} ) ( \pa + v_{1} ) ( \pa + v_{2} ) ( \pa - v_{2} ) ,
$$
which can be written as:
\be
L_{2} = ( \pa^2 + u + \varphi ) ( \pa^2 + u - \varphi ) , \label{hs}
\ee
with the Miura map:
\be
u = \frac{1}{2} ( v_{1x} - v_{2x} - v_{1}^{2} - v_{2}^{2} ), \quad
 \varphi = \frac{1}{2} ( v_{1x} + v_{2x} - v_{1}^{2} + v_{2}^{2} ) .
                  \label{miura2}
\ee
The operator (\ref{hs}) was found in \cite{f82-1} to be the Lax operator for
the Hirota-Satsuma system:
\be
 u_{t} = \frac{1}{2} u_{xxx} + 3 u u_{x} - 6 \varphi \varphi_{x}, \quad
       \varphi_{t} = -\varphi_{xxx} - 3 u \varphi_{x} ,    \label{hseqn}
\ee
which corresponds to the time evolution operator:
$$
M_{2} =  2 \pa^3 + 3 u \pa + \frac{3}{2} ( u_{x} - 2 \varphi_{x} )  .
$$

The isospectral flows of the general fourth order operator are
bi-hamiltonian, but the {\em first} Hamiltonian structure does not reduce to
the $2-$component submanifolds corresponding to our two operators.  As in the
case of our third order operators, {\em both} the $f,\,g$ and $u,\,\varphi$
hierarchies are related to the {\em same} modified hierarchy for the
functions $v_1$ and $v_2$, which has Hamiltonian form:
$$
\left( \begin{array}{c} v_1 \\ v_2 \end{array} \right)_t =
          \left( \begin{array}{cc}
                     - \pa & 0 \\ 0 & - \pa
          \end{array} \right)
               \left( \begin{array}{c} \delta_{v_1} h \\ \delta_{v_2} h
               \end{array} \right) .
$$
The Miura maps (\ref{miura1},\ref{miura2}) transport this structure
respectively onto:
$$
{\bf B}_{f} = \left( \begin{array}{cc}
                -2 \pa & - \pa^{2} - f \pa - f_{x} \\
     \pa^{2} - f \pa &
 \pa^{3} + ( 2 f_{x} + 2 g - f^{2} ) \pa + ( f_{xx} + g_{x} - f f_{x} )
             \end{array} \right)
$$
and:
$$
{\bf B}_{u} =  \left( \begin{array}{ll}
\frac{1}{2} \pa^3 + 2 u \pa + u_{x} & 2 \varphi \pa + \varphi_{x} \\
2 \varphi \pa + \varphi_{x} & \frac{1}{2} \pa^3 + 2 u \pa + u_{x}
                 \end{array} \right) .
$$
These are reductions of the general {\em second} Hamiltonian structure
associated with the general fourth order Lax operator.  With these, equations
(\ref{ckdv}) and (\ref{hseqn}) are respectively generated by
${\cal G} = -\frac{1}{8} ( f_{x}^{2} + f^{4} + 4 f g_{x} - 4 f^{2}
                                                         g - 4 g^{2} )$
and
${\cal H} = \frac{1}{2} u^{2} - \varphi^{2}$,
both of which are pulled back to the modified Hamiltonian
$h_{mod} = -\frac{1}{8} \left( v_{1x}^2 + v_{2x}^2 + v_1^4 + v_2^4 \right)
  -\frac{3}{4} \left( v_{1x} v_{2x} + v_2^2 v_{1x} - v_1^2 v_{2x}
                                                  - v_1^2 v_2^2 \right)$,
which evidently generates a coupled MKdV system.

\subsection{Stationary Flows and Related Quartic Potentials}

We now consider the stationary flows:
$$
{\bf B}_f \delta_{\bf f} {\cal G} = 0 \quad \hbox{and} \quad
                              {\bf B}_u \delta_{\bf u} {\cal H} = 0 ,
$$
which we show to be respectively related to generalisations of cases
(\ref{168}) and (\ref{161}) of the quartic potentials (\ref{quartic}).

\subsubsection{The $1:6:8$ Potential}

The stationary flow in the $(f,g)$ space is given by:
\bea
&&  ( 2 A + B_{x} + f B )_x = 0    \label{ft}   \\
&&  A_{xx} - f A_{x} + B_{xxx} + ( 2 f_{x} + 2 g - f^{2} ) B_{x}
                  + ( f_{xx} + g_{x} - f f_{x} ) B = 0 , \label{gt}
\eea
where, for the Hamiltonian $\cal G$,
\bea
A &=& \delta_f {\cal G} =
 \frac{1}{4} \left( f_{xx} - 2 g_x - 2 f^3 + 4 f g \right), \label{a} \\
B &=& \delta_g {\cal G} = \frac{1}{2} ( f_{x} + f^{2} + 2 g ) .  \label{b}
\eea
Equation (\ref{ft}) gives $A = \frac{1}{2} ( K - f B - B_x )$, where $K$ is a
constant of integration, and:

$$
B_{xxx} + 2 ( f_{x} + 2 g - \frac{1}{2} f^{2} ) B_{x}
                        + ( f_{xx} + 2 g_{x} - f f_{x} ) B = 0 ,
$$
which has solution $B = \alpha y^2$, where:
\be
y_{xx} + ( \frac{1}{2} f_{x} - \frac{1}{4} f^{2} + g ) y = L y^{-3}.
                                           \label{yxx}
\ee
Using (\ref{b}) we obtain a formula for $g$:
$$
g = \alpha y^{2} - \frac{1}{2} f^{2} - \frac{1}{2} f_{x}  ,
$$
which, from (\ref{a}), (\ref{yxx}) and the definition of $K$, gives:
\bea
&& f_{xx} - 2 f^{3} + 3 \alpha y^{2} f = K , \nn  \\
&& y_{xx} + ( \alpha y^{2} - \frac{3}{4} f^{2} ) y = L y^{-3} . \nn
\eea
When $\alpha = - \frac{1}{4}$ the above equations are Lagrangian with:
$$
{\cal L}_{(f,g)} = \frac{1}{2} ( f_{x}^{2} + y_{x}^{2} )
+ \frac{1}{16} ( y^{4} + 6 f^{2} y^{2} + 8 f^{4} )
                                  + K f - \frac{L}{2} y^{-2}.
$$
With canonical co-ordinates:
$$
Q_{1} = y ,\;\; Q_{2} = f,\;\; P_{1} = y_{x},\;\; P_{2} = f_{x},
$$
this gives the Hamiltonian:
\be
h_{(f,g)} = \frac{1}{2} ( P_{1}^{2} + P_{2}^{2} )
         -\frac{1}{16} ( Q_{1}^{4} + 6 Q_{1}^{2} Q_{2}^{2} + 8 Q_{2}^{4} )
       - K Q_{2} + \frac{L}{2} Q_{1}^{-2} , \label{ham168}
\ee
which is one of the integrable generalisations of case (\ref{168}) of
(\ref{quartic})  \cite{90-16}.

\subsubsection{The $1:6:1$ Potential}

The stationary flow in the $(u,\varphi)$ space is given by:
\bea
( \frac{1}{2} \pa^3 + 2 u \pa + u_{x} ) A
                      + ( 2 \varphi \pa + \varphi_{x} ) B &=& 0 , \nn \\
( 2 \varphi \pa + \varphi_{x} ) A
           + ( \frac{1}{2} \pa^3 + 2 u \pa + u_{x} ) B &=& 0 , \nn
\eea
where, for the simple Hamiltonian $\cal H$, we have
$A = \delta_u {\cal H} = u$ and $B = \delta_{\varphi} {\cal H} = - 2 \varphi$.
It is straightforward to find quadratic forms for $A$ and $B$:
$$
A = \alpha ( \psi^2 + \chi^2 ), \quad
                      B =  \alpha ( \chi^2 - \psi^2 ) ,
$$
where:
\bea
\psi_{xx} + ( u - \varphi ) \psi &=& l \psi^{-3} , \nn  \\
\chi_{xx} + ( u + \varphi ) \chi &=& k \chi^{-3} .   \nn
\eea
\br
The motivation for this representation of $A$ and $B$ is found in the squared
eigenfunction representation of $L_2 \theta = z \theta$, when written as a
$2 \times 2$, second order differential system.
\er
Upon substituting $u$ and $\varphi$ we get:
\bea
\psi_{xx} + \frac{1}{2} \alpha ( \psi^2 + 3 \chi^2 ) \psi
                                          &=& l \psi^{-3}, \nn  \\
\chi_{xx} + \frac{1}{2} \alpha ( 3 \psi^2 + \chi^2 ) \chi
                                         &=& k \chi^{-3} ,  \nn
\eea
which is Lagrangian with:
$$
{\cal L}= \frac{1}{2} ( \psi_{x}^{2} + \chi_{x}^{2} )
- \frac{1}{8} \alpha ( \psi^{4} + 6 \psi^{2} \chi^{2} + \chi^{4} )
        -\frac{1}{2} ( l \psi^{-2} + k \chi^{-2} ) .
$$
With the canonical co-ordinates:
$$
q_{1} = \psi , \;\; p_{1} = \psi_{x}, \;\;
                  q_{2} = \chi , \;\; p_{2} = \chi_{x} ,
$$
this gives (with $\alpha = 1$) the Hamiltonian:
\be
h_{u\varphi} = \frac{1}{2} ( p_{1}^{2} + p_{2}^{2} ) +
\frac{1}{8} ( q_{1}^{4} + 6 q_{1}^{2} q_{2}^{2} + q_{2}^{4} )
                    + \frac{1}{2} ( l q_{1}^{-2} + k q_{2}^{-2} ) ,
                               \label{ham161}
\ee
which is one of the integrable generalisations of case (\ref{161}) of
(\ref{quartic}) \cite{90-16}.

\subsection{The Lax Representations}

The Lax representations $L_{it} = [ M_i , L_i ]$ for the PDEs can be
re-written in zero curvature form:
$$
U_{it} - V_{it} + \left[ U_i , V_i \right] = 0 ,
$$
where:
$$
  U_{1}=\left( \begin{array}{cccc}
                        0& 1&0&0\\
                         -g&f&1&0\\
                         0&0&0&1\\
                     z& 0&-g-f_{x}&-f
                \end{array} \right)  ,  \quad
U_{2}=\left( \begin{array}{cccc}
                 0& 1&0&0\\
               \varphi-u&0&1&0\\
                 0&0&0&1\\
                   z& 0&-\varphi-u&0
\end{array} \right) ,
$$
and $V_i$ given by more complicated formulae.
The stationary flows are given by the Lax representations
$V_{ix} = [ U_i , V_i ]$.
which can be written in terms of $q_i , p_i$ and
$Q_i , P_i$.  Whilst $U_i$ are very simple, we need to use the equations
of motion generated by $h_{(f , g)}$ and $h_{(u , \varphi )}$ to eliminate
second and higher derivatives in $V_i$.  When written in this way the $V_i$
are the Lax matrices for the equations of motion and can be used to generate
the constants of motion.  They are given by:
\bea
V_{1} &=& \left( \begin{array}{cccc}
\frac{1}{8} Q_{2} Q_{1}^{2} + \frac{1}{4} Q_{1} P_{1} - \frac{1}{2} K
                         &- \frac{1}{4} Q_{1}^{2}&2 Q_{2}&2 \\
a_{21}&-\frac{1}{8} Q_{2} Q_{1}^{2} - \frac{1}{4} Q_{1} P_{1} - \frac{1}{2} K
                     &P_{2} + \frac{1}{4} Q_{1}^{2} + Q_{2}^{2} & 0 \\
            -2 Q_{2} z & 2 z & a_{33} &- \frac{1}{4} Q_{1}^{2} \\
z ( Q_{2}^{2} + \frac{1}{4} Q_{1}^{2} - P_{2} ) & 0 & a_{43} & a_{44}
                                        \end{array} \right) , \nn  \\
V_{2} &=& \left( \begin{array}{cccc}
           - 2 q_{1} p_{1} & 2 q_{1}^{2} & 0 & 2 \\
2 z - 2 p_{1}^{2} - 2 l q_1^{-2} & 2 q_1 p_1
                       &- ( q_1^2 + q_2^2 ) & 0 \\
0 & 2 z & - 2 q_2 p_2 & 2 q_2^2 \\
- z ( q_1^2 + q_2^2 ) & 0 & 2 z - 2 p_2^2 - 2 k q_2^{-2} & 2 q_2 p_2
           \end{array} \right) ,                            \nn
\eea
where:
\bea
a_{21} &=& \frac{1}{16} Q_1^2 Q_2^2 + \frac{L}{4Q_1^2}
       + \frac{1}{4} P_1^2 + \frac{1}{4} Q_1 Q_2 P_1 + 2 z , \nn  \\
a_{33} &=& - \frac{1}{8} Q_{2} Q_{1}^{2}
                   + \frac{1}{4} Q_{1} P_{1} + \frac{1}{2} K , \nn  \\
a_{43} &=& \frac{1}{16} Q_1^2 Q_2^2 + \frac{L}{4 Q_1^2}
          + \frac{1}{4} P_1^2 - \frac{1}{4} Q_1 Q_2 P_1 + 2 z , \nn  \\
a_{44} &=& \frac{1}{8} Q_2 Q_1^2 - \frac{1}{4} Q_1 P_1 + \frac{1}{2} K .
                                                         \nn
\eea
The first integrals are given by the characteristic equations:
\bea
\hbox{det}( V_1 - \mu I ) &=& \mu^4 + \frac{1}{8} (L-4 K^2) \mu^2
                                    + \hbox{det} V_1 ,
                                               \label{detv1}  \\
\hbox{det}( V_2 - \mu I ) &=& \mu^4 + 4 (k+l) \mu^2 + \hbox{det} V_2 ,
                                                \label{detv2}
\eea
where
\bea
\mbox{det}(V_{1}) &=&
-16 z^{3} - 8 h_{(f,g)} z^{2} - \frac{1}{4} k_{(f,g)} z
                    + \frac{1}{256} (L + 4 K^2)^2 , \nn   \\
\mbox{det} ( V_{2} ) &=& -16 z^{3} + 32 h_{(u,\varphi)} z^{2}
                     - 16 k_{(u,\varphi)} z + 16 lk .             \nn
\eea
Here $h_{(f,g)}$ and $h_{(u,\varphi)}$ are given by
(\ref{ham168},\ref{ham161}) and:
\bea
k_{(f,g)} &=& P_1^4 + \frac{1}{16} ( Q_2^2 Q_1^6 + Q_2^4 Q_1^4 )
- \frac{1}{4} ( Q_1^4 P_1^2 + Q_1^4 P_2^2 ) + \frac{1}{64} Q_1^8
+ P_1 P_2 Q_2 Q_1^3 - \frac{3}{2} Q_1^2 Q_2^2 P_1^2  \nn  \\
     &&    - \frac{1}{2} ( Q_1^4 Q_2 K + L Q_2^2 ) - 2 ( Q_1^2 K^2
                - \frac{P_1^2 L}{Q_1^2} )
- \frac{1}{4} Q_1^2 L + ( \frac{L^2}{Q_1^4} - Q_1^2 Q_2^3 K )  \nn  \\
 &&  + 4 ( Q_1 P_1 P_2 K - Q_2 P_1^2 K - \frac{ L K Q_2}{Q_1^2} ) . \nn \\
k_{(u,\varphi)} &=&
\left( p_1 p_2 + \frac{1}{2} ( q_1^3 q_2 + q_1 q_2^3 ) \right)^2
                + k q_1^2 + \frac{lk} {q_1^2 q_2^2}
+ \frac{k p_1^2}{q_2^2} + l q_2^2 + \frac{l p_2^2}{q_1^2} . \nn
\eea
The Lax equations for $V_i$ are known to linearise on the Jacobi variety of
the algebraic curves defined by (\ref{detv1}) and (\ref{detv2}), which gives
one of the standard integration procedures (see, for instance, Theorem 1,
page 67 of \cite{90-16})

\subsection{Gauge and Canonical Transformations}

The matrices $U_i,\, V_i$ (in the PDE variables) are related by gauge
transformation:
$$
U_2 = A U_1 A^{-1} + A_x A^{-1}, \quad V_2 = A V_1 A^{-1} + A_t A^{-1} ,
$$
with $A$ given by:
$$
A= \left( \begin{array}{cccc}
                     -v_{1} & 1 & 0 & 0 \\
                   -v_{1} v_{2} & v_{2} & 1 & 0 \\
                    0 & 0 & v_{2} & 1 \\
                     z & 0 & -v_{1} v_{2} & -v_{1}
                  \end{array} \right),
$$
whenever the Miura maps (\ref{miura1}) and (\ref{miura2}) are satisfied.
In terms of $q_i$ and their derivatives, the Miura maps can be
rearranged to give:
\bea
\frac{2}{3} ( v_{1x} - v_{1}^{2} ) &=&
      q_1^2 + \frac{1}{3} q_2^2 =
          -\frac{2 q_{2xx}}{3 q_2} + \frac{2}{3} k q_2^{-4} , \nn  \\
 \frac{2}{3} ( v_{2x} + v_2^2 ) &=&
                    q_2^2 + \frac{1}{3} q_1^2  =
       - \frac{2 q_{1xx}}{3 q_1} + \frac{2}{3} l q_1^{-4}.   \nn
\eea
We can then write $v_{1}$ and $v_{2}$ in terms of $q_{1}$ and $q_{2}$:
$$
v_{1} = - \frac{q_{2x}}{q_2} + \frac{a}{q_2^2}
        \quad \hbox{and} \quad
           v_{2} = \frac{q_{1x}}{q_1} + \frac{b}{q_1^2} ,
$$
where $a^2 = -k$ and $b^2 = -l$.  We can similarly write $v_i$ in terms of
$Q_1$ and $Q_2$:
$$
v_1 = \frac{1}{2} Q_{2} + \frac{Q_{1x}}{Q_1}
                       + \frac{1}{2} \frac{c}{Q_1^2},
                       \quad \hbox{and} \quad
v_2 = \frac{1}{2} Q_{2} - \frac{Q_{1x}}{Q_1}
                          -\frac{1}{2} \frac{c}{Q_1^2},
$$
where $c = 8 ( b - a )$.

We use these formulae, together with the Miura maps
(\ref{miura1},\ref{miura2}) and the definitions of the canonical coordinates,
to {\em explicitly} write down the canonical transformation:
\bea
Q_1 &=& 2 \Upsilon^{1 \over 2} , \qquad\qquad\qquad\qquad\quad
P_1 = \left( -\frac{p_1}{q_1}- \frac{p_2}{q_2}
          + \frac{a}{q_2^2} - \frac{b}{q_1^2} \right) \Upsilon^{1/2}
                               -\frac{c}{4} \Upsilon^{-1/2} , \nn  \\
Q_2 &=& \frac{p_1}{q_1} - \frac{p_2}{q_2}
                    + \frac{a}{q_2^2} + \frac{b}{q_1^2} , \qquad
P_2 = q_1^2 - q_2^2 + \frac{p_2^2}{q_2^2} - \frac{p_1^2}{q_1^2}
             -\frac{2 a p_2}{q_2^3} - \frac{2 b p_1}{q_1^3}
                           + \frac{a^2}{q_2^4} - \frac{b^2}{q_1^4} , \nn
\eea
and its inverse:
\bea
q_1 &=& \frac{1}{2} \Gamma_{+}^{1/2}  ,  \qquad
p_1 = \frac{1}{2}
      \left(
         \frac{1}{2} Q_2 - \frac{P_1}{Q_1} - \frac{c}{2 Q_1^2}
              \right)\Gamma_{+}^{1/2}  - 2 b \Gamma_{+}^{- 1/2} , \nn \\
q_2 &=& \frac{1}{2}\Gamma_{-}^{1/2} , \qquad
p_2 = -\frac{1}{2}
          \left(
             \frac{1}{2}Q_2 + \frac{P_1}{Q_1} + \frac{c}{2 Q_1^2}
                \right) \Gamma_{-}^{1/2}
                   + 2 a \Gamma_{-}^{- 1/2}  ,        \nn
\eea
where
\bea
\Upsilon &=&  q_1^2 + q_2^2 + \frac{2}{q_1 q_2}
                 \left( p_1 p_2 + \frac{b p_2}{q_1} - \frac{a p_1}{q_2}
                           -\frac{a b}{q_1 q_2} \right)  ,  \nn  \\
\Gamma_{\pm} &=& \pm 2 P_2 + Q_2^2 + \frac{1}{2} Q_1^2
               - \frac{4 P_1^2}{Q_1^2} \mp \frac{4 Q_2 P_1}{Q_1}
                     \mp \frac{2 c Q_2}{Q_1^2} -\frac{4 c P_1}{Q_1^3}
                            - \frac{c^2}{Q_1^4}  , \nn
\eea
The original parameters $k,l,K,L$ are related to $a$ and $b$ by:
$$
k = - a^2 , \; l = - b^2 , \; L = -16 (b-a)^2 , \; K = -2 (a+b).
$$
\br
When $a=b=0$, the case $1:6:1$ separates in Cartesian co-ordinates and the
canonical transformation between cases $1:6:1$ and $1:6:8$, constructed in
this paper, gives the separation co-ordinates for $1:6:8$ obtained in
\cite{94-1} by a Painlev\'e approach.  Some sort of Lax representation is
given in \cite{94-1}, but this involves the square root of $k_{(f,g)}$
and its derivatives, so does not give a clear derivation of the equations of
motion as a consequence of the Lax equations.
\er

\section{Conclusions}

The whole essence of our approach is to give a systematic connection between
completely integrable, finite dimensional Hamiltonian systems and integrable
(in the soliton theory sense) PDEs.  This result is, itself, interesting but,
more importantly, gives a systematic construction of a matrix Lax pair for
the finite dimensional system {\em and} (as a bonus) gives a straightforward
construction of canonical transformations (via gauge transformations in the
PDE framework).

We just presented the most interesting cases ((\ref{161}) and (\ref{168})) of
(generalisations of) Hamiltonian (\ref{quartic}) in this letter, but the
method is quite general.  Case (\ref{121}) can be obtained from the KdV
hierarchy, since it is just a special case of the Garnier system.  Case
(\ref{11216}) is a special case of a class studied in \cite{93-5}, associated
with energy dependent Schr\"odinger operators.  Case (\ref{11216}) can also
be obtained by considering a fifth order Lax operator \cite{baker}.

A more complete treatment will be presented elsewhere.

\subsection*{Acknowledgements}
SB was supported by an SERC (now EPSRC) studentship and VZE by the Royal
Society.  We thank both these organisations.


\end{document}